\documentclass[aps,twocolumn,secnumarabic,nobalancelastpage,footinbib, showpacs]{revtex4}

\usepackage{amsmath,amsfonts,amssymb}
\usepackage{wrapfig}
\usepackage{subfigure}
\usepackage{graphicx}
\usepackage{bbm}
\usepackage{graphics}
\usepackage{sidecap}
\usepackage{subfigure}

\def \beq {\begin{equation}}
\def \eeq {\end{equation}}
\def \ba {\begin{eqnarray}}
\def \ea {\end{eqnarray}}

\def\ket#1{\left| #1\right>}
\def\bra#1{\left< #1\right|}

\begin{document}
\title{Nonlinear Optics Quantum Computing with Circuit-QED}
\author{Prabin Adhikari$^1$, Mohammad Hafezi$^{1,2}$, J.M. Taylor$^{1,2}$}
\affiliation{$^1$Joint Quantum Institute, University of Maryland, College Park}
\affiliation{$^2$National Institute of Standards and Technology, Gaithersburg, Maryland}

\begin{abstract}
One approach to quantum information processing is to use photons as quantum bits and rely on linear optical elements for most operations. However, some optical nonlinearity is necessary to enable universal quantum computing \cite{KLM, Kok,Nielsen, Browne}. Here, we suggest a circuit-QED approach to nonlinear optics quantum computing in the microwave regime, including a deterministic two-photon phase gate. Our specific example uses a hybrid quantum system comprising a LC resonator coupled to a superconducting flux qubit to implement a nonlinear coupling. Compared to the self-Kerr nonlinearity, we find that our approach has improved tolerance to noise in the qubit while maintaining fast operation.
\end{abstract}

\pacs{03.67.Lx, 84.40.Dc, 85.25.Cp}

\maketitle

Linear optics quantum computing (LOQC) has proven to be one of the conceptually simplest approaches to building novel quantum states and proving the possibility of quantum information processing. This approach relies on the robustness of linear optical elements, but implicitly requires an optical nonlinearity \cite{KLM, Kok, Nielsen, Browne}. Unfortunately, progress towards larger scale systems remains challenging due to the limits to optical nonlinearities, such as the measurement of single photons \cite{Buller, Schuster}.

In this letter we suggest recent advances in circuit-QED in which optical and atomic-like systems in the microwave domain are explored for their novel quantum properties, provides a new paradigm for quantum computing with photons \cite{Blais1, Wallraff, Aumentado}, which, in contrast to LOQC, is deterministic. Specifically, using superconducting nonlinearities in the form of Josephson junctions and the related quantum devices such as flux and phase qubits \cite{Martinis1, Blais2}, key elements of our approach have been realized: the creation of microwave photon Fock states \cite{Aumentado, Eichler, Houck, HofHeinz}, controllable beam splitters \cite{Xiao, Aumentado}, and single microwave photon detection \cite{Johnson, Romero}. In many cases, the photons stored in a transmission line-based resonator or inductor-capacitor resonator have much better coherence times than the attached superconducting qubits \cite{Yin, Nakamura, Paik}. This suggests that the main impediment to photon based quantum computing is the realization of appropriate photon nonlinearities to enable two-qubit gates like two-photon phase gates, which are sufficient for universal quantum computation \cite{KLM, DiVincenzo}. 

\begin{figure}[t]
\centering 
\includegraphics[width=8cm]{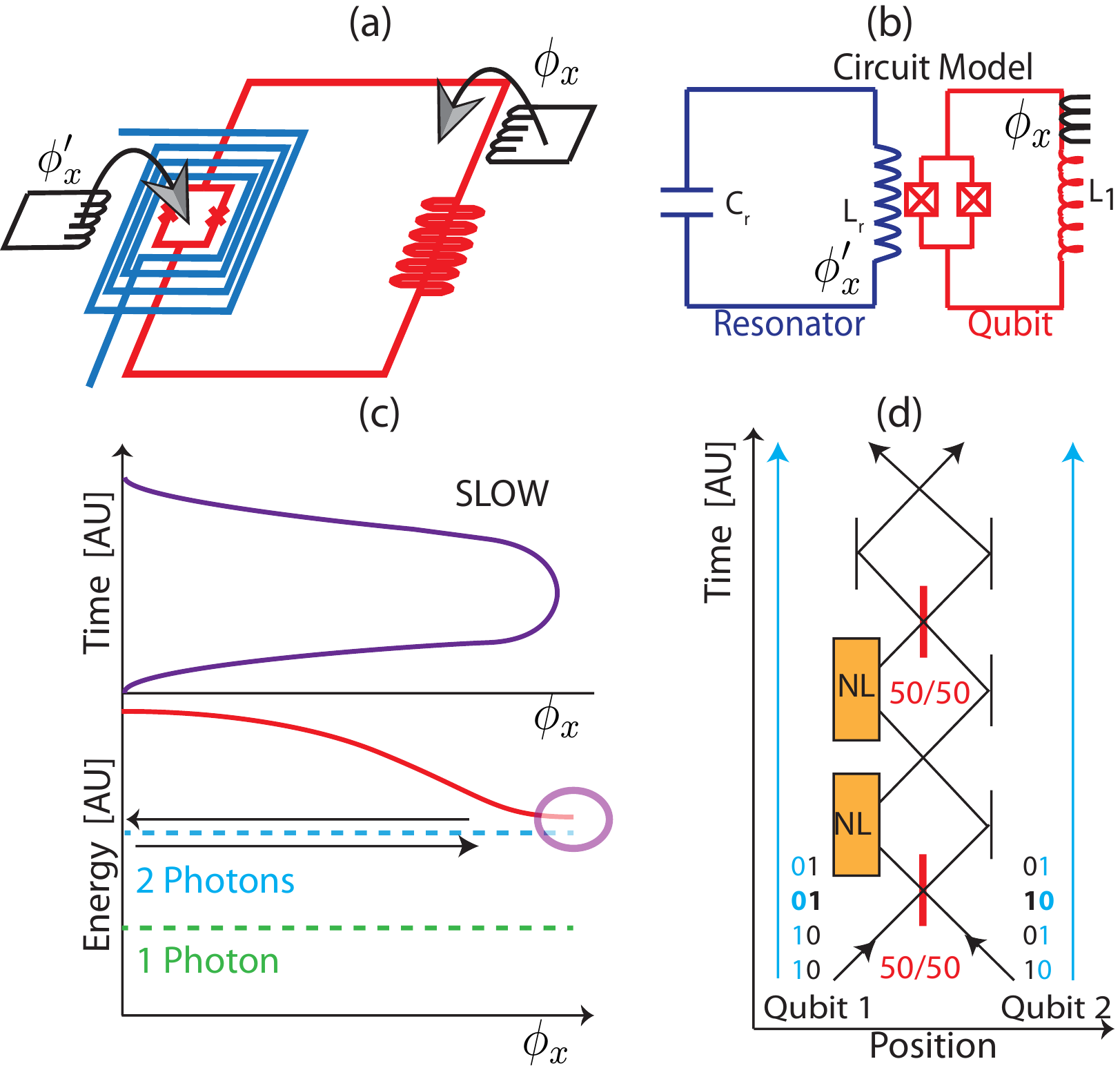} 
\caption{(a) Implementation of a high-impedance coiled resonator (blue) coupled to a dc SQUID (red) with an inductive outer loop. The flux bias lines are in black. (b) A simple circuit model of our physical implementation. (c) Bottom: Energy levels of the coupled system with a sizeable two-photon coupling. Top: The suggested flux bias pulse $\phi_{x}$ to implement the nonlinear phase shift; a fast but adiabatic sweep and then a very slow variation of the pulse near the avoided crossing. (d) Use of two nonlinear phase shifters, combined with 50/50 beamsplitters, leads to a deterministic two photon phase gate using dual rail logic. The two photons in the dual rail basis $\ket{0}_{L} \ket{1}_{L} = \ket{0 1}_{1} \ket{1 0}_{2}$ of the two qubits become bunched into a single mode after passing through the first beam splitter, and then receive a $\pi$ phase from the nonlinear phase shifter. Storage cavities (not shown in (b)) are blue lines.}
\label{1}
\end{figure} 

The key element of a two-photon phase gate is a two-photon nonlinear phase shifter. It imparts a $\pi$ phase on any state consisting of two photons, while leaving single photon and vacuum states unaffected. A deterministic approach to achieve such photon nonlinearity is based on the Kerr effect \cite{Yin, Nemoto, Hoi, Milburn}. In the context of circuit-QED, in  Ref.\cite{Milburn}, a four level N scheme using a coplanar waveguide resonator and a Cooper pair box is used to arrange for EIT \cite{Fleischhauer} to generate large Kerr nonlinearities. In this Letter we take a different approach to photon nonlinearity. We explore the possibility of using a dc SQUID \cite{Clarke} to implement a nonlinear coupling between qubit and resonator, which, through an adiabatic scheme, enables a high fidelity, deterministic two-photon nonlinear phase shift in the microwave domain. Along with the nonlinearity, we envision using dynamically controlled cavity coupling to implement a 50/50 beam splitter operation to construct a two-photon phase gate using so-called dual rail photon qubits \cite{Aumentado, Chuang}, in which the logical basis \{$\ket{0}_{L} = \ket{01}, \ket{1}_{L} = \ket{10} \}$ corresponds to the existence of a single photon in one of two resonator modes (Figure~\ref{1}d). Our approach takes best advantage of the relatively long coherence times for microwave photons in resonators, and couples only virtually to superconducting quantum bit devices, minimizing noise and loss due to errors in such devices. When combined with the aforementioned techniques for Fock state generation and detection, along with dynamically controlled beam splitters, this provides the final element for nonlinear optics quantum computing in the microwave domain. 

We now outline our approach. We consider photons stored in a high-impedance microwave resonator \cite{Teufel} coupled inductively with strength $0<\chi<1$ to a flux superconducting qubit (SQ) in a dc SQUID configuration (Figure~\ref{1}a). The resonator loops around the dc SQUID which results in a nonlinear cosine dependent interaction between the resonator and qubit. In this configuration, we get an effective coupling of the form $V \sim E_J \cos(\hat{\phi} + \phi_{x}') \cos \hat{\phi}_{r}$, where an external flux $\phi_{x}' \equiv 2 \pi \chi \Phi_{x}' /\Phi_{0}$ is applied to the resonator which consequently threads the smaller loop of the dc SQUID, $\Phi_{0}$ being the superconducting flux quantum. The qubit phase variable and the resonator flux are denoted by $\hat{\phi}$ and $\hat{\phi}_{r} = 2 \pi \hat{\Phi}_{r} / \Phi_{0}$ respectively. For $\phi_{x}' \sim \pi/2$, we see immediately a nonlinear coupling between the qubit and resonator: $V \sim E_J \hat{\phi} \hat{\phi}_{r}^2$, where two resonator photons can be annihilated to produce one qubit excitation, analogous to parametric up conversion in $\chi^{(2)}$ systems. This causes the two-photon state of the resonator to couple to the first excited state of the SQ with strength $g_{2}$ (Figure~\ref{1}c). In essence, in this region, the two-photon state with detuning $\delta$ from the qubit, becomes slightly qubit-like and acquires some nonlinearity. However, the single-photon state, inspite of its coupling to the first excitation of the SQ with strength $g_{1}$, remains mostly photon-like because it is far detuned by $\Delta$ from this qubit excitation. At the end of the procedure, this leads to an additional phase for the two-photon initial state. The coupling  of the two-photon state to other modes arises via linear coupling at $O(g_{1})$ and is assumed to be far detuned. 

The noise in the SQ, with a decay rate $\gamma$ of its first excited state, may slightly limit our nonlinear phase shift operation. Although the overall system will mostly be in the photon-like regime with decay rate $\kappa$, there will be an additional probability for it to decay due to its coupling to the lossy qubit. In the limit where $|\delta| \gg |g_{2}|$ and $|\Delta| \gg |g_{1}|$ with $|\Delta| > |\delta|$, the two-photon nonlinearity  goes like $g_{2}^2 / \delta$, and the two-photon state decays approximately at a rate $\gamma g_{1}^2/ \Delta^{2}+\gamma g_{2}^2 / \delta^2$. Thus, the losses due to the qubit go like $\gamma/\delta$ provided we allow $g_{1}$ to become close to $g_{2}$, which is possible by controlling $\phi_{x}'$. Hence, at large detuning, we will then be limited only by $\kappa$. In contrast, a Kerr nonlinearity scales like $g_{1}^4 / \delta^3$ and the noise scales like $\gamma g_{1}^2 / \delta^2$, leading to more loss due to the qubit for large detuning. 

We now examine a detailed model to support these qualitative arguments. In our case, the second resonator is not coupled to a SQ and is not shown; we focus on the dynamics of the first resonator, which is coupled. The quantum Hamiltonian of the system is $H=T+V$ \cite{Devoret}.
\ba
T &=& \frac{\hat{q}_{r}^2}{2 C_{r}} + \frac{\hat{q}^2}{2 C_{J}} - \frac{\chi}{2 C_{r}} \hat{q} \hat{q}_{r},\\
V &=& \left(\frac{\Phi_{0}}{2 \pi}\right)^2 \frac{\hat{\phi}_{r}^2}{2L_{r}} - E_{J}[ \cos (\hat{\phi} + \chi \hat{\phi}_{r}+\phi_{x}' ) + \cos \hat{\phi}] \label{Potential} \nonumber   \\
& & + \frac{E_{L}}{2} (\hat{\phi}+\phi_{x})^2 . \label{V}
\ea 
In addition to $\phi_{x}'$, an external flux bias $\phi_{x} = 2 \pi \Phi_{x} / \Phi_{0}$ is applied to the outer inductor loop attached to the squid. The canonical coordinates of the qubit satisfy $[\hat{\phi},\hat{N}]=i$, where $\hat{N} = \hat{q} (2 e)^{-1}$ is the number of Cooper pairs in the junctions. The operators $\hat{\Phi}_{r}$ and $\hat{q}_{r}$ represent quantum fluctuations in flux and charge of the resonator satisfying $[\hat{\Phi}_{r}, \hat{q}_{r}]= i \hbar$, and $\chi$ is the fraction of the flux $\hat{\Phi}_{r}$ threading the squid loop. This inductive coupling causes the effective capacitances of the resonator and qubit to be modified, and we denote these modified values by $C_{r}$ and $C_{J}$ respectively. $E_{J}$ is the Josephson energy of each junction, while $E_{L}= \Phi_{0}^2/(4 \pi^2 L_{1})$ represents the inductive energy of the qubit due to the bigger loop.  We can also define an effective charging energy of the junction to be $E_{C} = (2e)^2 C_{J}^{-1}$. We introduce another dimensionless parameter $\mu = 2 \pi \Phi_{0}^{-1} \Phi_{r}^0$, where $ \Phi_{r}^0 =\sqrt{L_{r} \omega \hbar/2}$ is the width of quantum fluctuations in the resonator flux.  In terms of the quantum of conductance $G_{0} = 2e^2/h$ and the characteristic impedance of the resonator  $Z = (L_{r}/C_{r})^{1/2}$, we can write $\mu = \sqrt{2 \pi G_{0} Z}$. Since $\mu \ll 1$, we can expand $V$ in powers of $\chi \hat{\phi}_{r} \propto \mu$. Performing the expansion to second order, we get $H = H_{r} + H_{q} + V_{I}$ with
\ba
H_{r} &=&  \frac{\hat{q}_{r}^2}{2 C_{r}}+\frac{\hat{\Phi}_{r}^2}{2L_{r}} \label{Numerics1} \\
H_{q} &=&  \frac{\hat{q}^2}{2 C_{J}}-2 E_{J} \cos \left[\frac{\phi_{x}'}{2}\right]  \cos \left[\hat{\phi}+\frac{\phi_{x}'}{2} \right] + \frac{E_{L}}{2}(\hat{\phi}+\phi_{x})^2 \nonumber \\
V_{I} &=& \chi E_{J} [\hat{\phi}_{r}  \sin (\hat{\phi}+\phi_{x}') + \frac{\chi  \hat{\phi}_{r}^2}{2}  \cos (\hat{\phi}+\phi_{x}')] - \frac{\chi \hat{q} \hat{q}_{r} }{2 C_{r}} \nonumber
\ea
corresponding to the resonator, qubit, and interaction terms respectively. We remark that asymmetry in the two Josephson junctions leads to additional terms, but our general linearization approach described below remains valid, and provides qualitatively similar results.

The equations \eqref{Numerics1} can be used to solve this system numerically. However, in the regime where $E_{L} \gg E_{J}$ we can get some analytical results. First, we linearize the potential in \eqref{Potential} around the classical values of the resonator reduced flux $\phi_{cl}$ and the qubit phase $\beta_{cl}=-\phi_{x}+f$, with quantum fluctuations $\hat{\varphi}_{r}$ and $\hat{\varphi}$ around them. Any nonlinearity can then be treated as a perturbation. Note the following functions of $\phi_{x}$ and $\phi_{x}'$.
\ba
\phi_{cl} &=& \frac{E_{J} L_{r} \chi \sin (\phi_{x}-\phi_{x}')}{\Phi_{0}^2/(2 \pi)^2 +  E_{J} L_{r} \chi^2 \cos (\phi_{x}-\phi_{x}')}, \\
f &\equiv& \frac{E_{J} [\sin \phi_{x}+ \sin(\phi_{x} - \phi_{x}' - \chi \phi_{cl})]}{E_{L} + E_{J} [  \cos \phi_{x} +   \cos(\phi_{x} -\phi_{x}'- \chi \phi_{cl})]}, \\
r &\equiv& \sin \beta_{cl}; \ s \equiv \sin \left[\beta_{cl}+\phi_{x}' + \chi \phi_{cl} \right], \\
t &\equiv& \cos \beta_{cl}; \ u \equiv \cos \left[\beta_{cl} + \phi_{x}' + \chi \phi_{cl} \right].
\ea
With the effective inductance of resonator $\tilde{L}_{r}^{-1} = L_{r}^{-1} + (2 \pi/\Phi_{0})^2 E_{J} \chi^2 u$ and  $\hat{\varPhi}_{r} = \Phi_{0}/(2\pi) \hat{\varphi}_{r}$, the resonator and qubit Hamiltonians can now be written as
\ba
H_{r} = \frac{\hat{q}_{r}^2}{2 C_{r}} +  \frac{\hat{\varPhi}_{r}^2}{2 \tilde{L}_{r}}; H_{q} = \frac{\hat{q}^2}{2 C_{J}} + \frac{E_{L} + E_{J} (t+u)}{2} \hat{\varphi}^2
\ea
with respective frequencies $\omega = (\tilde{L}_{r} C_{r})^{-1/2}$ and $\omega_{q} = \sqrt{\omega_{C} \left[ \omega_{L} + \omega_{J} ( t + u )   \right]}$. We see immediately that changing the external fluxes changes these frequencies, and hence, the qubit-resonator detuning $\Delta = \omega_{q}- \omega$. Introducing creation and annihilation operators for the resonator and qubit satisfying $[\hat{a},\hat{a}^{\dagger}]=1=[\hat{b},\hat{b}^{\dagger}]$ with
\ba
\hat{\varphi} &=& \sqrt{\frac{\omega_{C}}{2 \omega_{q}}}(\hat{b}+\hat{b}^{\dagger}) ; \  \hat{N} = -i \sqrt{\frac{\omega_{q}}{2 \omega_{C} }}(\hat{b}-\hat{b}^{\dagger}), \\
\hat{\varPhi}_{r} &=& \sqrt{\frac{\tilde{L}_{r} \omega \hbar}{2}} (\hat{a}+\hat{a}^{\dagger})  ; \  \hat{q}_{r} =  -i \sqrt{\frac{\hbar}{2 \tilde{L}_{r} \omega}}  (\hat{a}-\hat{a}^{\dagger}). \nonumber
\ea
the resonator and qubit Hamiltonians become $H_{r} = \omega \hat{a}^{\dagger} \hat{a}$ and $H_{q} = \omega_{q} \hat{b}^{\dagger} \hat{b}$. When the qubit is not linearized as in \eqref{Numerics1}, the potential energy terms can be written as 
\ba
V_{1} &=& \eta_{1} (\hat{a} + \hat{a}^{\dagger}) \sin (\hat{\phi}+\phi_{x}'),  \label{V1} \\
V_{2} &=& \eta_{2} (\hat{a} + \hat{a}^{\dagger})^2 \cos (\hat{\phi}+\phi_{x}'), \label{V2} \\
V_{3} &=& i \eta_{3} (\hat{a} - \hat{a}^{\dagger}) \hat{N} \label{V3},
\ea
with coupling coefficients $\eta_{1} = \chi E_{J} \mu$, $\eta_{2} = \eta_{1}^2 / (2 E_{J})$, and $\eta_{3} = (\eta_{1} \hbar \omega) / ( 2 E_{J})$. The potential that is of relevance is $V_{2}$ from which the nonlinear coupling term $g_{2}$ is seen to be 
\ba
g_{2} = \sqrt{2} \eta_{2} \bra{0_{q}} \cos (\hat{\phi}+\phi_{x}') \ket{1_{q}},
\ea
where the matrix element is between the ground and first excited qubit states. The size of $g_{2}$ is important for the success of the nonlinear phase shift protocol. For a given value of $\eta_{1}$, the Josephson energy $E_{J}$ of each junction cannot be made too large as this will suppress $\eta_{2}$ and  $g_{2}$. Hence, it is desirable to operate the qubit in the flux regime where $E_{J} \leq 10 E_{C}$. The linear coupling coefficient $\eta_{1}$ depends on the characteristic impedance $Z$ of the LC circuit implicit in the parameter $\mu$. Therefore, we have to implement a resonator with high $Z$ to make $g_{2}$ larger. 

After linearization, the flux dependent  linear Hamiltonian of the system can be written as
\ba
H_{L} &=& H_{r} + H_{q}  - \frac{\chi}{2 C_{r}} \hat{q} \hat{q}_{r} + \chi E_{J} u \ \hat{\varphi}_{r} \ \hat{\varphi}.
\ea
We neglect all higher order nonlinear terms and only consider the perturbative $\chi^{(2)}$ type nonlinearity given by,
\ba
V_{2} &=& -E_{J} \chi^2 s/2  \ \hat{\varphi} \ \hat{\varphi}_{r}^2.
\ea
We can make the rotating wave approximation and write $H_{L}$ in terms of creation and annihilation operators as
\ba
H_{L} &=& \omega \hat{a}^{\dagger} \hat{a} + \omega_{q} \hat{b}^{\dagger} \hat{b} + g_{1} (\hat{a} \hat{b}^{\dagger} + \hat{a}^{\dagger} \hat{b}) \label{HLin},
\ea
where the linear coupling $g_{1}$ is given by
\ba
g_{1} &=&  \eta_{1} u \sqrt{\frac{\omega_{C}}{2 \omega_{q}}} - \eta_{3} \sqrt{\frac{\omega_{q}}{2 \omega_{C}}}.
\ea

To diagonalize $H_{L}$ we define new operators $\hat{c}$ and $\hat{d}$ as
\ba
\hat{a} &=& \mu_{1} \hat{c} + \nu_{1} \hat{d} ; \hat{b} = \mu_{2} \hat{c} + \nu_{2} \hat{d} \label{Transform},
\ea
such that $[\hat{c},\hat{c}^{\dagger}] = 1 = [\hat{d},\hat{d}^{\dagger}]$ and $[\hat{c},\hat{d}^{\dagger}] = 0 = [\hat{c},\hat{d}]$. This requires the conditions
\ba
|\mu_{1}|^2+|\nu_{1}|^2 = 1 = |\mu_{2}|^2+|\nu_{2}|^2  ; \mu_{1} \mu_{2}^{\star} + \nu_{1} \nu_{2}^{\star} = 0.  \label{Canonical}
\ea
The parametrization $\mu_{1} = \cos \theta$, $\nu_{1} = - \sin \theta$, $\mu_{2} = \sin \theta$, $\nu_{2}= \cos \theta$ satisfies the constraints \eqref{Canonical}. Substituting the relations \eqref{Transform} into $H_{L}$ and setting the diagonal terms to zero, we get a normal mode Hamiltonian $H_{N} = \Omega_{1} \hat{c}^{\dagger} \hat{c} + \Omega_{2} \hat{d}^{\dagger} \hat{d}$ with energies $\Omega_{1,2} = \omega + \Delta/2 \left(1 \mp  \sqrt{1 + 4 g_{1}^2/\Delta^2} \right)$.
We assume the detuning $\Delta > 0$.  The bare basis states of the resonator-qubit system will be denoted by $\ket{n} \otimes \ket{q} \equiv \ket{n  \  q }$, where the first  and second labels refer to the quantum numbers of the resonator and qubit respectively.  The relevant eigenstates of the Hamiltonian in the new basis are number excitations of the operators $\hat{c}^{\dagger} \hat{c}$ and $\hat{d}^{\dagger} \hat{d}$. Denoting these kets as $\ket{\bar{C} \bar{D}}$, we can write down three important eigenstates with energies $\Omega_{1}$, $\Omega_{2}$, and $2 \Omega_{1}$. They are
\ba
\ket{\bar{1} \bar{0}} &=& \cos \theta \ket{1 0} + \sin \theta \ket{0 1}, \\
\ket{\bar{0} \bar{1}} &=& -\sin \theta \ket{1 0} + \cos \theta \ket{0 1}, \nonumber \\
\ket{\bar{2} \bar{0}} &=& \cos^2 \theta \ket{2 0}+\sqrt{2} \cos \theta \sin \theta \ket{1 1}+\sin^2 \theta \ket{02}. \nonumber 
\ea
The parameter $\theta$ satisfies  $\tan 2\theta = -2 g_{1} \Delta^{-1}$. For $\Delta \gg |g_{1}|$, $\ket{\bar{1} \bar{0}}\rightarrow \ket{10}$, $\ket{\bar{0} \bar{1}} \rightarrow \ket{01}$, $\ket{\bar{2} \bar{0}} \rightarrow \ket{20}$,\ $\Omega_{1} \rightarrow \omega$, and $\Omega_{2} \rightarrow \omega_{q}$. 

The nonlinearity in our model couples the states $\ket{\bar{2} \bar{0}}$ and $\ket{\bar{0} \bar{1}}$ leading to a sizeable avoided crossing in Figure~\ref{1}c between the two-photon and qubit levels. In terms of the normal mode operators
\ba
V_{2} =  \eta_{2}' [\cos^2 \theta \sin \theta \hat{c}^{\dagger 2} \hat{c} - \cos^3 \theta  \hat{c}^{\dagger 2} \hat{d} - 2 \cos \theta \sin^2 \theta \hat{c}^{\dagger}\hat{c} \hat{d}^{\dagger}  \nonumber \\
+ 2 \cos^2 \theta \sin \theta \hat{c}^{\dagger} \hat{d}^{\dagger} \hat{d}   + \sin^3 \theta \hat{c}  \hat{d}^{\dagger 2} - \sin^2\theta \cos \theta \hat{d}^{\dagger 2} \hat{d}] + \rm{HC}, \nonumber
\ea
where $\eta_{2} ' =  \eta_{2} s/\sqrt{2}$. Then the overall Hamiltonian of interest becomes $H = \Omega_{1} \hat{c}^{\dagger} \hat{c} + \Omega_{2} \hat{d}^{\dagger} \hat{d} + V_{2}$. 

Working in the truncated subspace spanned by the states $\{\ket{0} \equiv \ket{\bar{0} \bar{0}}, \ket{a} \equiv \ket{\bar{1} \bar{0}}, \ket{b} \equiv \ket{\bar{2} \bar{0}}, \ket{c} \equiv \ket{\bar{0} \bar{1}} \}$, we write the Hamiltonian as   $H=H_{0}+V$ where $H_{0} = \Omega_{1} \ket{a}\bra{a} + 2 \Omega_{1} \ket{b}\bra{b}+ \Omega_{2} \ket{c}\bra{c}$ and the coupling $V= \lambda_{1} (\ket{a}\bra{b}+\ket{b}\bra{a}) +\lambda_{2} (\ket{b}\bra{c}+\ket{c}\bra{b})$. The parameters $\lambda_{1} = \sqrt{2} \eta_{2}'  \cos^2 \theta \sin \theta \equiv r_{1} \eta_{2}'$ and $\lambda_{2} = -\sqrt{2} \eta_{2}'  \cos^3 \theta \equiv r_{2} \eta_{2}'$. We can adiabatically eliminate the state $\ket{a}$ to find an effective Hamiltonian 
\ba
H_{e} &=&  \left[\Omega_{1} - \frac{r_{1}^2 \eta_{2}'^2}{\Omega_{1}} \right] \ket{a}\bra{a} + \left[2 \Omega_{1} + \frac{r_{1}^2 \eta_{2}'^2}{\Omega_{1}} \right] \ket{b}\bra{b} \nonumber \\
&+&  \Omega_{2} \ket{c} \bra{c} + r_{2} \eta_{2}' (\ket{b} \bra{c} + \ket{c} \bra{b}).
\ea

We can use this Hamiltonian to calculate the two-photon nonlinearity $N_{l}$. For $|\delta'| \equiv |\Omega_{2} - 2 \Omega_{1}| \gg |\eta_{2}' r_{2}|$, we have
\ba
N_{l} &=& - \frac{(\eta_{2}' r_{2})^2}{\delta'} \equiv -\frac{g_{2}^2}{\delta'} \approx -\frac{g_{2}^2}{\delta}.
\ea
where we have associated the nonlinear coupling $g_{2}$ with $\eta_{2}' r_{2}$. The nonlinear phase-shift protocol requires initializing the system in the states $\ket{10} \approx \ket{\bar{1} \bar{0}}$ and $\ket{20} \approx \ket{\bar{2} \bar{0}}$ with errors that go like  $g_{1}^2 / \Delta^2$. Then the external fluxes $\phi_{x}$ and $\phi_{x}'$ are varied adiabatically so that the state $\ket{\bar{2} \bar{0}}$ becomes slightly qubit-like, mostly because of $\ket{11}$. After accumulating the desired phase, the process is reversed to retrieve the photons. For some integer $n$, we require for a total time $\tau_{g}$, $\int_{0}^{\tau_{g}}  N_{l} (t)  dt = (2 n + 1) \pi$. The final outcome is then $\frac{1}{\sqrt{3}} (\ket{00} + \ket{10} + \ket{20} ) \rightarrow \frac{1}{\sqrt{3}} (\ket{00} + \ket{10} - \ket{20} )$. 

In addition to our analytical model, we also diagonalize the Hamiltonian of the system numerically by working in the tensor product space $H = H_{r} \otimes H_{q}$ of the resonator and qubit using the Hamiltonian in \eqref{Numerics1}. The basis states in the resonator space are the number excitations $\ket{n}$. The qubit space is written in the basis of qubit wavefunctions $\psi_{q}(\phi) = \bra{\phi} q \rangle$. We let $\hbar=1$ and choose $\omega_{C}/ (2 \pi) = 1 \ \rm{GHz}$, $\omega_{J}/(2 \pi)=5 \ \rm{GHz}$, $\omega_{L}=3 \omega_{J}$, and $\omega/(2 \pi)= 2.225 \ \rm{GHz}$. The characteristic impedance $Z \approx 449 \ \rm{\Omega}$. We choose a $\chi =0.17$, representing an easily achievable mutual inductance, from which follow $\eta_{1} /(2 \pi)= 400 \ \rm{MHz}$, $\eta_{2}/(2 \pi)=16 \ \rm{MHz}$, and $\eta_{3}/(2 \pi)=89 \ \rm{MHz}$. 
 
\begin{figure}[h]
\centering 
\includegraphics[width=8.6cm]{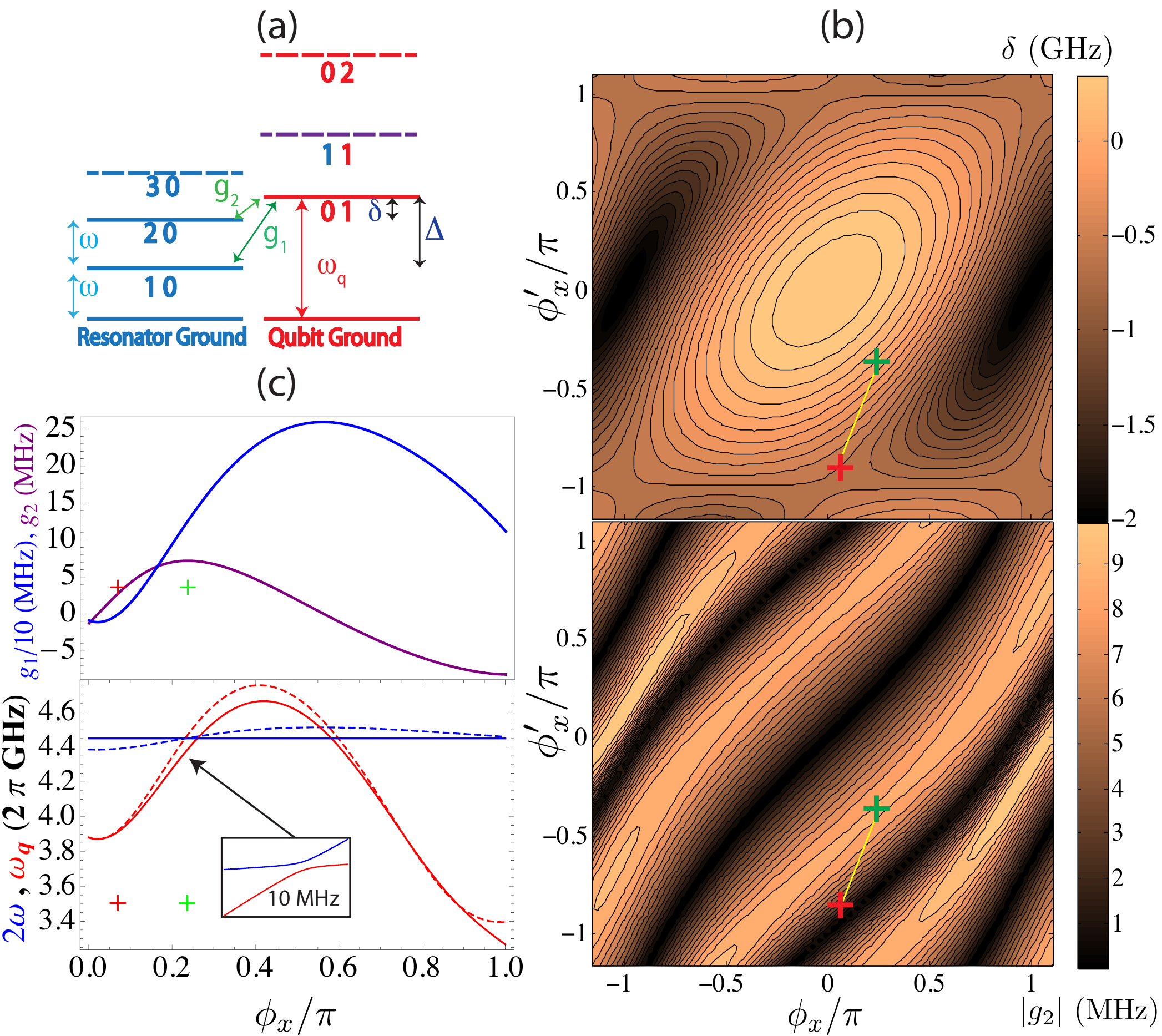} 
\caption{(a) Schematic of the system bare energy levels and couplings. (b) Contour plot of detuning $\delta$ and $|g_{2}|$ with the on and off points marked in green and red. The on point is chosen such that the $g_{2}$ is maximized. (c) Top: The coupling $g_{1}/10$ and $g_{2}$, with the on and off fluxes shown. Bottom: The bare frequencies $2 \omega/(2\pi)$ and $\omega_{q}/(2 \pi)$ obtained from the analytical model (dashed) and numerical results (solid). The overall qubit-resonator interaction leads to a splitting of approximately $10 \ \rm{MHz}$. }
\label{2}
\end{figure} 

Now we discuss the effect of loss on our gate. Since throughout the operation of the gate the system remains photon-like, loss is dominated by the cavity decay at a rate $\kappa$. For the photon-like state $\ket{\bar{2} \bar{0}}$, there are two other decay channels due to the cavity-qubit coupling. The linear coupling $g_{1}$ in the limit $\Delta \gg |g_{1}|$ leads to $\gamma_{1} \equiv \gamma g_{1}^2/\Delta^2=\gamma g_{1}^2/(\delta+ \omega)^2$. Similarly the nonlinear coupling leads to $\gamma_{2} \equiv \gamma g_{2}^2 / \delta^2$ for  $|\delta| \gg |g_{2}|$. Including the cavity decay rate $\kappa$, the total decay rate of the two-photon-like state becomes $\Gamma(\delta) = \kappa + \gamma_{1}+\gamma_{2}$. 

\begin{figure}[htbp]
\includegraphics[width=\linewidth]{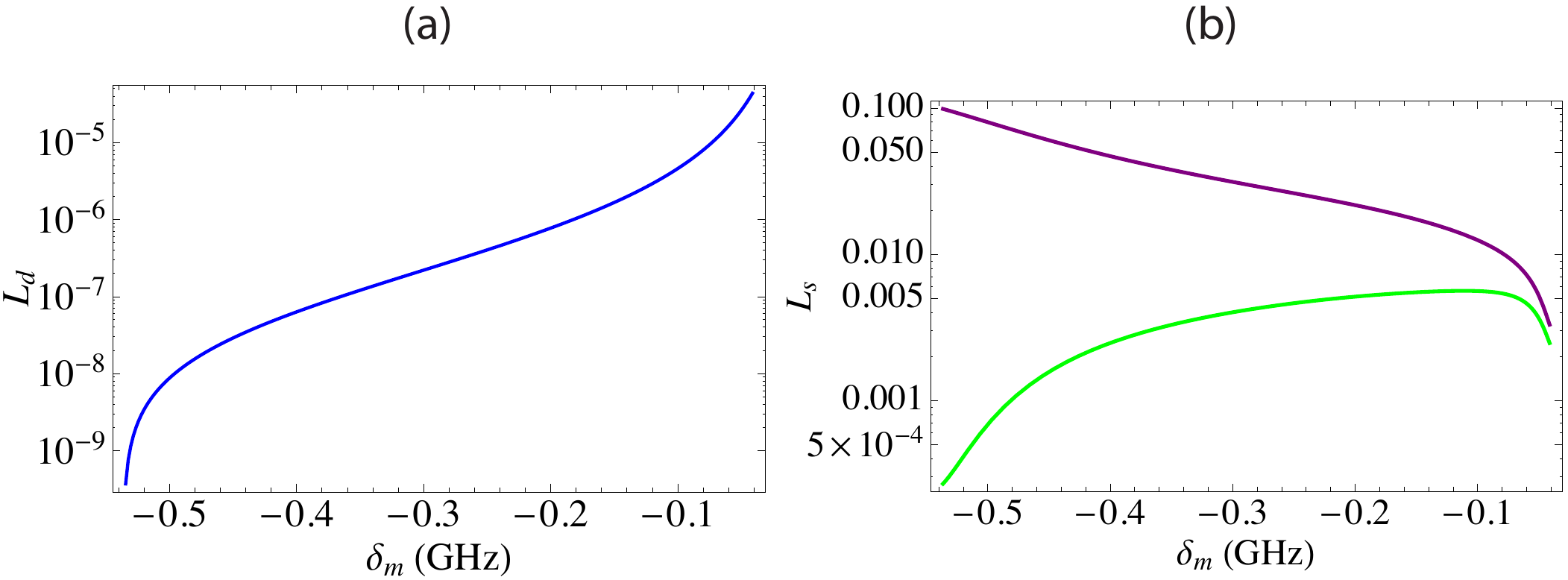}
\caption{(a) A plot of the dimensionless dynamic loss $L_{d}$ for $\kappa = 1 \ \rm{KHz}$, $\gamma = 100 \kappa$ and $\epsilon^2 = 0.01$. The detuning $-536 \ \rm{MHz} \leq \delta_{m} \leq  -41\ \rm{MHz}$. (b) The static loss $L_{s}$ in purple, and the static loss without the effect of the cavity decay rate $\kappa$ in green.}
\label{3}
\end{figure}

Assuming $g_{2}$ is time independent for simplicity, adiabaticity requires $g_{2}^2 |\dot{\delta}|^2  (\delta^2 + 4 g_{2}^2)^{-3} \ll 1$ .  We can set this equal to some  $\epsilon^2 \ll 1$ and solve for 
\ba
\tau_{h}(\delta_{m}) &=& - \frac{1}{\epsilon} \int_{\delta_{i}}^{\delta_{m}} \frac{|g_{2}|}{(\delta^2+4 g_{2}^2)^{\frac{3}{2}}} d\delta,
\ea
which is the time taken to go from $|\delta_{i}| \gg |g_{2}|$ at $t=0$ to smaller values of detuning with a minimum $\delta_{m}$. The total dynamic loss during the process is given by
\ba
L_{d}(\delta_{m}) = \frac{2}{\epsilon} \int_{\delta_{m}}^{\delta_{i}} \Gamma(\delta) \frac{|g_{2}|}{(\delta^2 + 4g_{2}^2)^{\frac{3}{2}}} d \delta.
\ea
When the detuning is held at $\delta_{m}$ for a time $\tau_{s} = \pi  \delta_{m} / g_{2}^2$, the static loss  $L_{s}(\delta_{m}) = \tau_{s} \Gamma(\delta_{m})$. Thus,
\ba
L_{s}(\delta_{m}) =  \pi \left[\frac{\kappa \delta_{m}}{g_{2}^2} + \frac{\gamma \delta_{m}}{(\delta_{m}+\omega)^2} \left(\frac{g_{1}}{g_{2}}\right)^2 + \frac{\gamma}{\delta_{m}} \right] \label{Ls},
\ea
and the total time of the protocol is $\tau_{g} = 2 \tau_{h} + \tau_{s}$. Assuming $\delta_{m} \ll \omega$, $L_{s}(\delta_{m})$ is minimized when $\delta_{m} \approx g_{2} \sqrt{\gamma / \kappa}$. However, the on-off ratio of the photon nonlinearity goes like $|\delta_{i}/\delta_{m}|$, and a value of $\delta_{m}$ that makes this ratio at least a hundred is desirable. For $\delta \sim \omega$, we can make $g_{1}\approx g_{2}$ so that $L_{s}(\delta) < \kappa \delta / g_{2}^2 + 2 \gamma / \delta$. In this regime $L_{s}$ is limited by $\kappa$, as  can be verified from Figure~\ref{3}b. Thus, we optimize our protocol so that the loss $L = L_{d} + L_{s} \ll 1$. We note that our protection is only against qubit noise and loss, and comes at the cost of increased reliance on the cavity quality factor. 

The protocol might also be limited by dephasing of the qubit due to flux noise \cite{Martinis2, Bialczak, Kakuyanagi}. The average slopes of the single and two-photon energy levels with respect to the reduced flux $\phi_{x}$ are approximately $50 \ \rm{MHz}$ and $100 \ \rm{MHz}$ respectively, while the slope of the qubit energy level is at most $1 \ \rm{GHz}$ for the parameters chosen. However, the exact loss due to dephasing depends on the flux noise amplitude \cite{Steffen, Anton}.

In conclusion, we have demonstrated that by appropriately tuning the two control fluxes, the nonlinear coupling enables a two-photon nonlinear phase shift operation with loss at large detuning limited only by the cavity quality factor. This is highly desirable compared to the self-Kerr nonlinearity which leads to more loss due to the qubit for large detunings. Furthermore, our approach may be adaptable to recent ultra-high quality factor resonators enabling nonlinear optics quantum computing in a fully engineered system \cite{Paik}.

The authors wish to thank E. Tiesinga, J. Aumentado, A. Blais, and S. Girvin for helpful discussions. This research was supported by the US Army Research Office MURI award W911NF0910406 and the NSF through the Physics Frontier Center at the Joint Quantum Institute.

\end{document}